\documentclass[conference]{IEEEtran}
\IEEEoverridecommandlockouts
\usepackage{fontawesome}
\usepackage{booktabs}
\usepackage{xcolor}
\usepackage{todonotes}
\usepackage{paralist}

\usepackage{amsthm}

\usepackage{graphicx}

\def\BibTeX{{\rm B\kern-.05em{\sc i\kern-.025em b}\kern-.08em
    T\kern-.1667em\lower.7ex\hbox{E}\kern-.125emX}}

\begin{document}


\title{Towards a Systematic Engineering of \\ 
	Industrial Domain-Specific Languages}

\author{
	\IEEEauthorblockN{Rohit Gupta\IEEEauthorrefmark{1},
		Sieglinde Kranz\IEEEauthorrefmark{1},
		Nikolaus Regnat\IEEEauthorrefmark{1},
		Bernhard Rumpe\IEEEauthorrefmark{2} and
		Andreas Wortmann\IEEEauthorrefmark{3}}
	\IEEEauthorblockA{\IEEEauthorrefmark{1}
		Siemens AG, Munich, Germany\\
		Email: rg.gupta@siemens.com, sieglinde.kranz@siemens.com, nikolaus.regnat@siemens.com}
	\IEEEauthorblockA{\IEEEauthorrefmark{2}Software Engineering, RWTH Aachen University, Aachen, Germany\\
		Email: rumpe@se-rwth.de}
	\IEEEauthorblockA{\IEEEauthorrefmark{3}Institute for Control Engineering of Machine Tools and Manufacturing Units, University of Stuttgart, Stuttgart, Germany\\
		Email: andreas@wortmann.ac}\\[-5.0ex]}

\maketitle

\begin{abstract}



%
%

Domain-Specific Languages (DSLs) help practitioners in contributing
solutions to challenges of specific domains.
The efficient development of user-friendly DSLs suitable for industrial
practitioners with little expertise in modelling still is challenging.
%
%
For such practitioners, who often do not model on a daily basis, there is a need to foster reduction of repetitive modelling tasks and providing simplified visual representations of DSL parts.
For industrial language engineers, there is no methodical support for
providing such guidelines or documentation as part of reusable language
modules.
%
%
Previous research either addresses the reuse of languages or guidelines for
modelling.
For the efficient industrial deployment of DSLs, their combination is
essential:
the efficient engineering of DSLs from reusable modules that feature
integrated documentation and guidelines for industrial practitioners.
%
%
To solve these challenges, we propose a systematic approach for the
industrial engineering of DSLs based on the concept of reusable DSL Building
Blocks, which rests on several years
of experience in the industrial engineering of DSLs and their deployment to
various organizations.
%
%
We investigated our approach via focus group methods consisting of five
participants from industry and research qualitatively.
%
%
Ultimately, DSL Building Blocks support industrial language engineers in developing
better usable DSLs and industrial practitioners in more efficiently 
achieving their modelling.
\end{abstract}

\begin{IEEEkeywords}
	Domain-Specific Languages, Model-Based Systems Engineering, Industrial Language Engineering
\end{IEEEkeywords}
\vspace{-0.75em}
\maketitle

\section{Introduction}
\vspace{-0.75em}

There is a conceptual gap~\cite{FR07} in the systems engineering domain between the
expertise of participating domain experts (biologists, chemists, mechanical
engineers, etc.) and the challenges of systems engineering.
Consequently, with the advance of various systems engineering domains from
documents to models, we are seeing a shift in the way modelling is
introduced at early stages of the systems engineering processes.
The ubiquitous General-Purpose Languages (GPLs) used for software
development present difficulties in system modelling \cite{mci/Proper2020} as
they focus on technical implementation details, aggravate analysing the
systems under development holistically, and prevent domain experts from
contributing solutions directly.
Domain-Specific Languages (DSLs)~\cite{DBLP:books/daglib/0034522} instead
aim to reduce the gap by being aimed at a particular domain, supporting
domain-specific abstractions, and are better accessible to analysis and
synthesis of systems and their parts.
In the context of this paper, we use the term \textit{developer} to refer to industrial language engineers and \textit{user} to refer to industrial practitioners and domain experts.




Various graphical DSLs have been developed to support modelling in different
domains, such as MATLAB Simulink~\cite{Cha15} or SysML \cite{FMS14}.
Yet, these are still overly generic and do not reflect domain concepts.
However, systematically developing truly domain specific languages, e.g., a
systems engineering language for the Italian railway
system~\cite{356ac9c6461040a79c84fa1ebf6815e0}, that captures this
particular domain's terminology (syntax), rules (well-formedness
constraints), and meaning (semantics) is complicated.
Reusing encapsulated DSL parts systematically can facilitate engineering new
DSLs and ultimately foster truly domain-specific systems modelling.
Additionally, the deployment of such DSLs to their users can be challenging.
Often, the users the DSLs are developed for, model quite rarely,
maybe once a week or less. Another challenge of graphical DSLs is to effectively represent DSL elements visually \cite{10.1109/TSE.2009.67, nielsen} that improves usability heuristics and aids users in deriving hints to the meaning of such elements with the use of icons, colours and appearances.
Hence, despite employing domain terminology, concepts, rules, and meaning,
modelling with many DSLs can be less effective than expected.
Industrial DSL engineering, therefore, needs to consider that users are perhaps
modelling less often than expected and integrate modelling support and
usability considerations into reusable language components, the DSL Building
Blocks.

Addressing the above challenges is essential to achieving modelling goals
effectively. Therefore, this paper presents an approach, DSL Building
Blocks, which supports developers in building better usable graphical DSLs more efficiently and helps users achieve modelling more efficiently.
\begin{figure}[h]
	\centering
	\includegraphics[width=\linewidth]{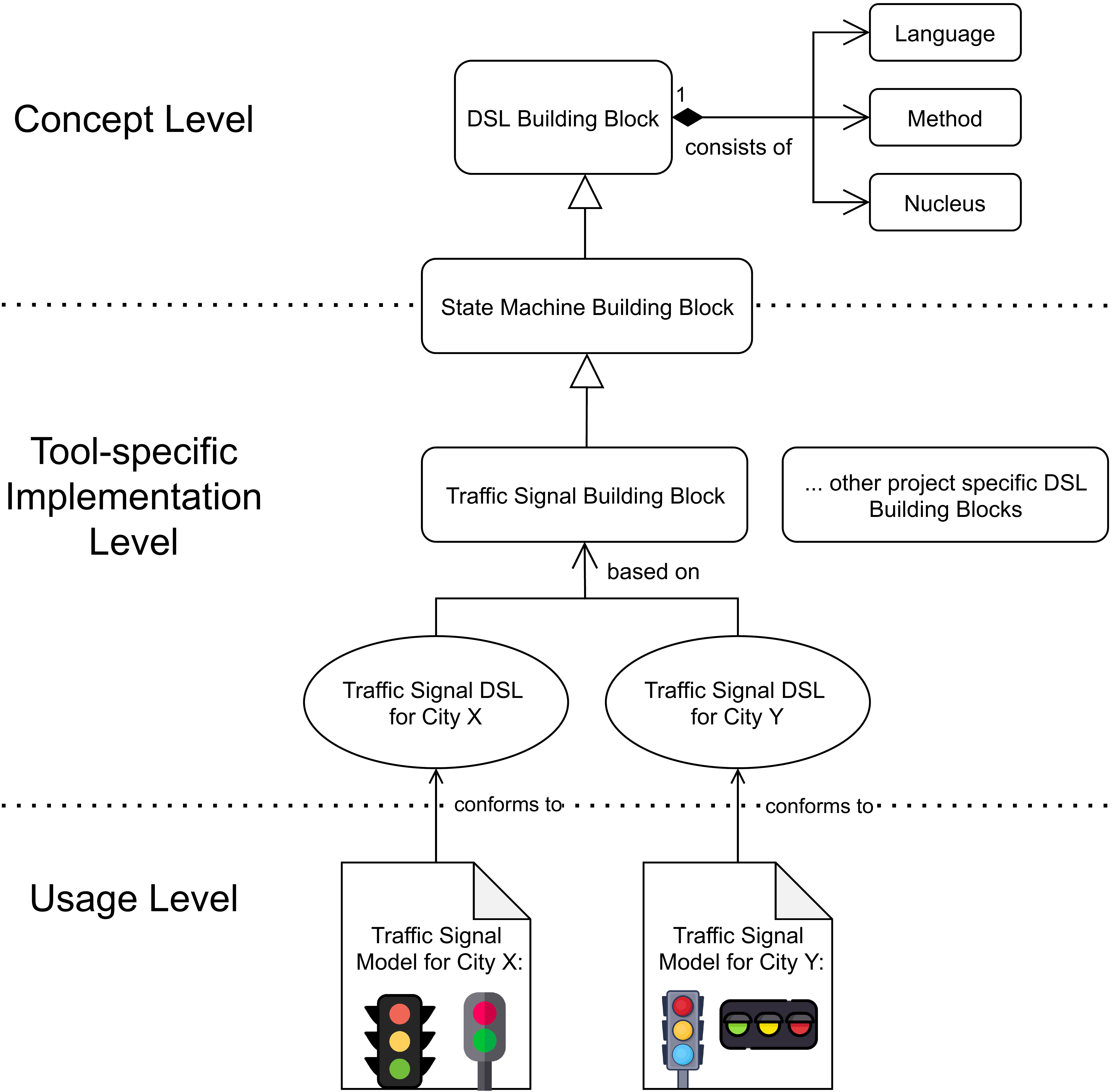}
	\vspace*{-1.75em}
	\caption{A conceptual model for the development and usage of a graphical DSL describing the different levels in the development process including defining the DSL Building Block consisting of the method, language and nucleus at the concept level. Different traffic signals layouts for City X and City Y are depicted at the usage level for different cities.}
	\label{fig:dslbb_structure}
	\vspace{-2.05em}
\end{figure}
Our approach, summarized in Fig. \ref{fig:dslbb_structure}, separates the concerns of industrial engineering and deployment
of DSLs along three different levels that relate to different skill sets and 
activities: 
\begin{inparaenum}[(1)]
\item Concept level: in this level, developers define three parts: (i) the language, where abstract syntax, graphical concrete syntax and the translational semantic mapping is defined for developers; (ii) the method, where certain constraints and methodical steps are described for users to help achieve intended modelling goals; and (iii) the nucleus, where visual representations and notations on DSL elements are described by developers to help users relate to commonly used visual designs; 
\item Tool-specific implementation level: in this level, developers realize the concepts described in the concept level by developing the DSL Building Blocks and the DSLs using a graphical modelling tool;
\item Usage level: this is the level where users understand the methodical steps, documentation and visual representations of DSL elements to achieve their modelling goals.
\end{inparaenum}
Leveraging this separation enables a systematic development of graphical DSL for developers who can reuse common parts for similarly structured DSLs. Further, it is beneficial to users as they can follow methodical steps and derive meaning from visual representation of DSL elements simplifying their modelling experience and efficiently reaching their modelling goals.
Our focus on this paper is on the concept level, as developers could use different graphical modelling tools to build their DSLs. By defining DSL constraints, steps to reach a modelling goal, language syntaxes and nucleus nuances, we separate the language, method and nucleus of a DSL Building Block enabling developers to re-use these parts during the  development of similarly structured graphical DSLs.


The concept of DSL Building Blocks is supported with a running example of a
state machine and investigated through a case study and a qualitative focus
group research~\cite{krueger2009focus}.
Overall, our approach builds on years of experience in developing graphical DSLs for various industrial projects. By separating the concerns of industrial DSL engineering across different levels, developers achieve reuse of encapsulated parts in a systematic way. Effective visual representation of DSL elements and a structured guidance to achieving modelling goals is beneficial for users in optimizing their modelling experience.


In the remainder, \ref{bg} provides some background and related work, before
\ref{md} presents our approach with a running example.
Afterward, \ref{cs} describes a case study based on the example and
\ref{eval} describes the evaluation.
Finally, \ref{discussion} provides a discussion of the approach and \ref{con} concludes the paper.

\vspace{-.7em}
\section{Background}
\label{bg}
\vspace{-.6em}

DSLs are software languages and software languages are software
too~\cite{FGD+10}.
Hence, their engineering is subject to the usual challenges of software
engineering in addition to considering multiple (meta)languages to define
the language's constituents.
Generally, a software language consists of~\cite{CGR09,CBCR15}:
\begin{inparaenum}[(1)]
	\item an abstract syntax that defines the structure of its models, e.g., in 
	form of grammars~\cite{HR17} or metamodels~\cite{CBW17};
	\item a concrete syntax that defines how the models are presented, e.g., 
	graphical~\cite{DCB+15}, textual~\cite{Bet16}, or projectional~\cite{Cam14}; and
	\item semantics, in the sense of meaning~\cite{HR04}, often realized through 
	model-to-text~\cite{Bet16} or model-to-model transformations~\cite{JFB+08}.
\end{inparaenum}

To address this complexity, the research area of Software Language
Engineering (SLE)~\cite{Kle08,HRW18} has emerged and with it, language
workbenches~\cite{ESV+13}, specialized tools for the creation of software
languages.
Many of these provide advanced language engineering support, such as the
generation of debuggers~\cite{DCB+15}, editors~\cite{Bet16}, or reusable
language modules~\cite{HR17} from abstract syntax descriptions. 

Yet, methodologies for systematic SLE in the large are rare.
And where studies detail how industrial graphical DSLs are implemented
\cite{Mendez-AcunaGDC16, TolvanenK05}, the
reported methodologies are highly specific to individual departments and
require repeated, time-consuming effort in language engineering. Either central research units, such as in Siemens Technology, must become more common in developing truly domain-specific or users in companies not having such units, such as Siemens Healthineers, must be trained with software and language engineering. 
Moreover, the language engineering and reuse methods proposed so far focus
solely on technical improvements, such as explicit language
interfaces~\cite{BPR+20}, merging of language parts~\cite{DCB+15}, or
language types~\cite{SJ07}.
Usability of languages and language parts, in the form of modelling
documentation, guidance, or automated modelling assistants as part of graphical DSL development are still missing.

In our literature review, we found two aspects that are
often neglected: a systematic approach to developing graphical DSLs that fosters re-usability of common language elements combining guidelines for modelling and ways to achieve efficient user experience (UX) for users of such DSLs.
Despite the efforts in previous research \cite{KKP+09, CzechMP20}, there is a lot to consider when building DSLs. We have not come across a methodology for developing graphical DSLs that takes into consideration how
to reach a certain modelling goal and ways to improve the UX for users.
One possible explanation could be that modelling goals vary for different
projects and usability heuristics are not considered part of the language
definition.
In our experience of developing a diverse set of graphical DSLs for
industrial projects, there is a need for a methodology that can be
described independent of a specific implementation or graphical modelling tool that eventually benefits
users.


\vspace{-.5em}
\section{DSL Building Blocks}
\label{md}
\vspace{-.25em}

In this section, we introduce the approach and terminology of a DSL Building
Block with the help of a running example of a state machine. A state machine
consists of \textit{states} (the information of a system at an
instant), \textit{transitions} (translations from a source to a target
state), and \textit{triggers} (actions or inputs that hold to enable a
transitions).
\textit{Initial} states describe where the state machine begins its
execution or gets reset to.
\textit{Final} states describe where the state machine ends its execution.
All other states are \textit{intermediate} states.

\vspace{-0.6em}
\subsection{DSL Building Block Structure}
\vspace{-.35em}



Our approach, summarized in Fig. \ref{fig:dslbb_structure} as a conceptual model, starts from a classical domain-driven approach where users define the business requirements such as number of states and transitions in a state machine. The systematic approach to developing graphical DSLs is based on domain-driven design \cite{evans2004ddd}, allowing developers to create graphical DSLs aimed at better modelling software and system architectures. These business requirements, specific to each project, are first translated into a \textit{DSL Building Block} that primarily consists of three parts: 
(1) the \textit{language}, which defines the abstract syntax, graphical concrete syntax and semantics of the language;
(2) the \textit{method}, which describes how to reach a modelling goal; and 
(3) the \textit{nucleus}, which describes the visual representations and notations for better UX on model elements. 
In our example, defining these parts leads to a structured definition of a state machine DSL Building Block, which is then extended to specific industrial examples such as different traffic signalling systems, an oven or a heater, or systems that require sequential control logic. 
The outcome of this developmental approach is to prevent the reinvention of the same method, language and nucleus parts of similar graphical DSLs more than once.

\begin{figure}[h]
	\centering
	\includegraphics[width=\linewidth]{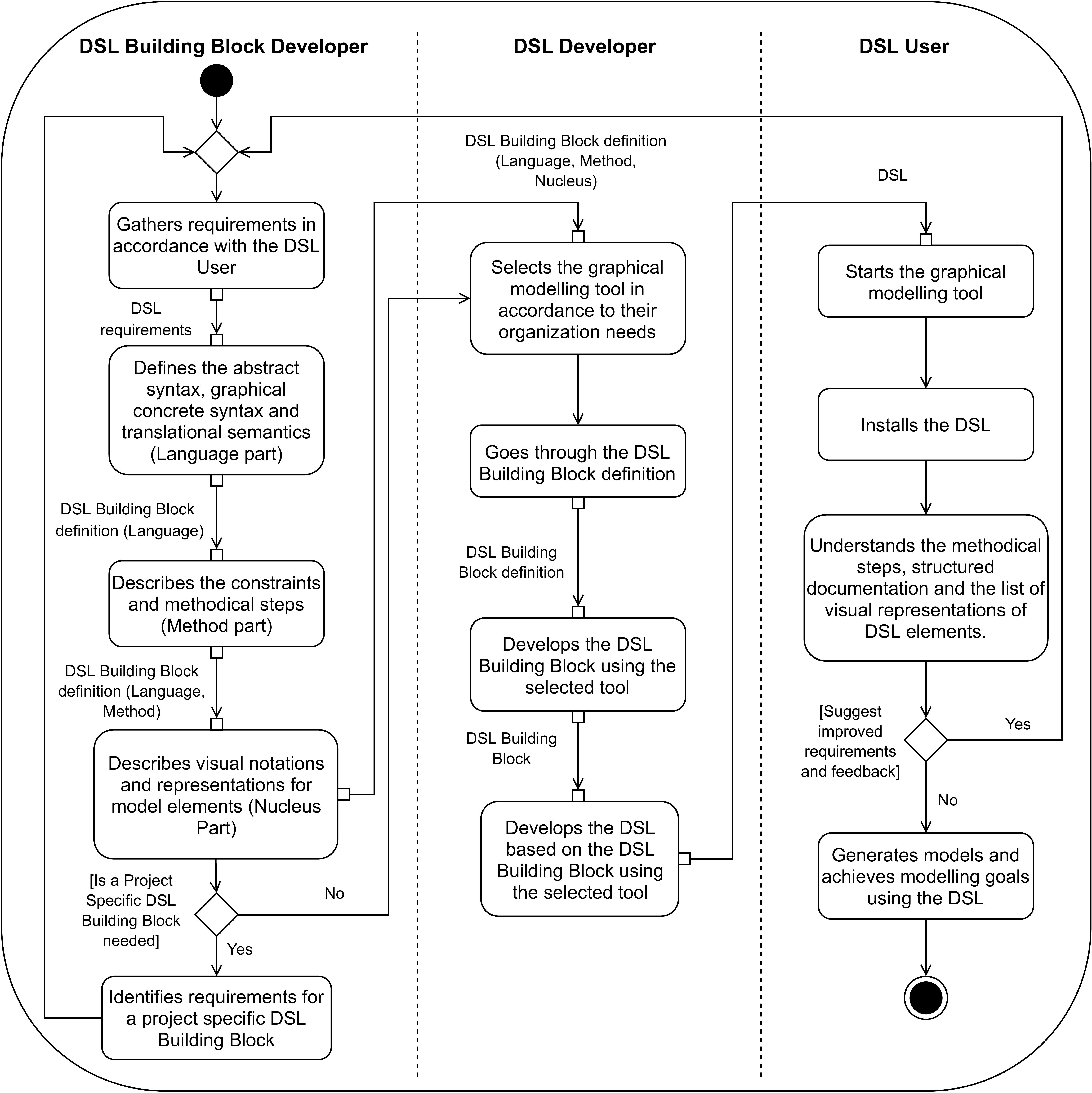}
	\vspace*{-1.75em}
	\caption{An activity diagram describing the tasks and activities of the three actors in the DSL Building Block approach.}
	\label{fig:actors_ad}
	\vspace{-1em}
\end{figure}

The development and usage of a graphical DSL based on DSL Building Blocks is segregated into three logical levels and is performed by three actors whose tasks and activities are described in Fig. \ref{fig:actors_ad}. 
At the \textit{concept} level, a \textit{DSL Building Block Developer}, who is a modelling expert   
with additional expertise in UX, and language engineering (such as key expert engineers in software and systems research units at Siemens), describes the constraints and method of use of the DSL, the abstract syntax, the graphical concrete syntax along with a structured documentation of model elements and the nucleus consisting of visual notations and representations of the model elements in accordance with the requirements specified by users. As well as defining the parts generically, such as for a state machine, they also identify and define project specific DSL Building Block requirements, such as for a traffic signal system.
The \textit{tool-specific implementation} level is where a \textit{DSL Developer}, equipped with sufficient programming skills needed for building graphical DSLs, selects the relevant graphical modelling tool in accordance to their organization and develops the DSL Building Block and its corresponding DSL. 
The \textit{usage} level is where users understand the steps, documentation and visual representations of DSL elements to ultimately achieve their modelling goals on the selected graphical modelling tool. This is performed by a \textit{DSL User}, who does not need to be proficient in UX, modelling or programming, but possess expert knowledge in their respective domains. They also provide subsequent feedback and improved requirements to DSL Building Block Developers corresponding to agile principles of immediate feedback and continuous integration \cite{Beck00}. As DSL Users represent a diverse set of users, it is important for a DSL Building Block Developer to describe the parts of a DSL Building Block in a meaningful and easy to understand manner using appropriate documentation and visual designs and notations belonging to a particular domain. While developers could also possess domain expertise, they are rather uncommon in Siemens.
\vspace{-.5em}

We now describe each of the parts based on the state machine example. It is worth mentioning that we represent the state machine model elements within square brackets such as [State: X], [Transition: y]. X and y are the names or an identification number for the respective model element in a derived DSL Building Block. For a state machine DSL Building Block the main model elements are defined as {[State]}, {[Transition]} and {[Trigger]}.
\vspace{-.70em}
\subsection{The Language}
\begin{figure}[h]
	\centering
	\includegraphics[totalheight=5.5cm]{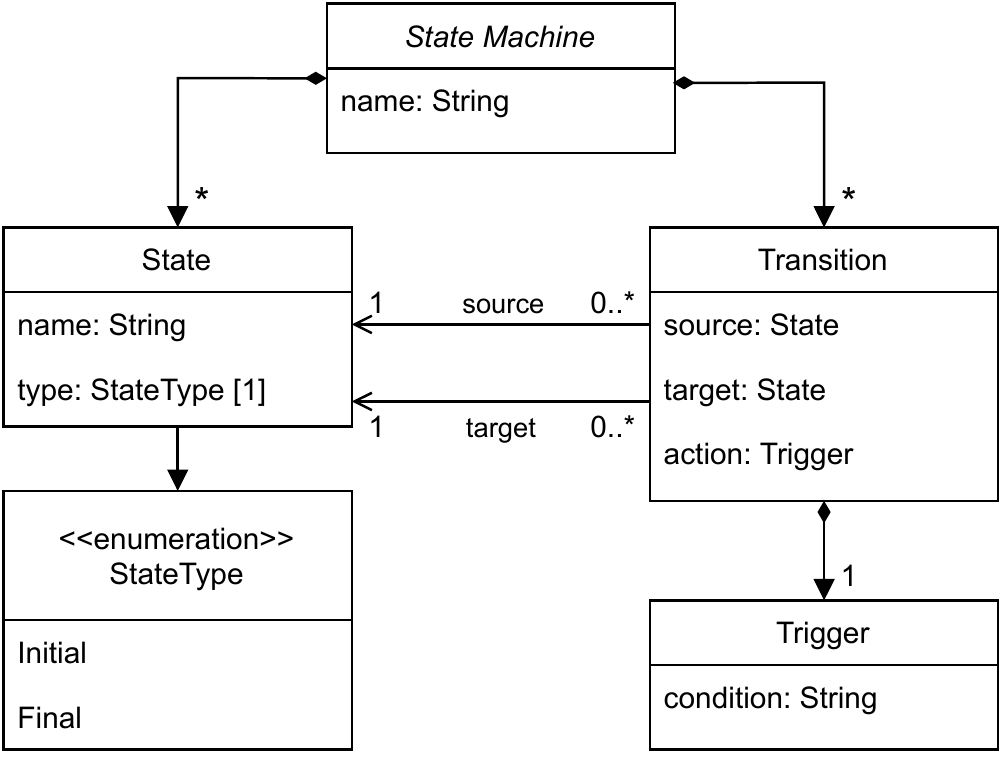}
	\vspace*{-1em}
	\caption{A class diagram showing the abstract syntax of a state machine DSL Building Block as well as the relationships between them.}
	\label{fig:dslbb_2_sm_cd}
	\vspace{-1.7em}
\end{figure}

\vspace{-0.35em}
The \textit{language} part of the DSL Building Block describes the abstract syntax, the graphical concrete syntax and the translational semantic mapping for the language. Fig. \ref{fig:dslbb_2_sm_cd} shows the abstract syntax of a state machine describing the necessary concepts and structure of the language. The model elements {[State Machine]}, {[State]}, {[Transition]}, {[Trigger]} and the {[StateType]} are defined along with the list of attributes. In our example, the method list is left empty for the class diagram, but can be modified later to allow for specific business requirements. The DSL Building Block Developer initially defines a default graphical form without any specific visualization properties, for each model element as part of the concrete syntax. E.g., a state is defined to be a rectangle corresponding to a UML object and a transition is defined to be a straight line. These model elements are visually defined and improved later in the nucleus to complete a better graphical concrete syntax aimed for a more visually understandable and easy to use DSL. The structured documentation from the defined model elements to the semantic domain essentially provides meanings to those model elements. For the state machine, a structured documentation of the model elements and their attributes along with detailed description is shown in Table \ref{tab:semantic-mapping-sm}, for DSL Users to easily understand and relate each model element.
\vspace{-1.5em}

\begin{table*}[h]
	\centering
	\caption{The structured documentation of syntax elements for a state machine DSL Building Block.}
	\vspace{-1.5em}
	\begin{center}
		\begin{tabular}{@{}|l|l|l|@{}}
			\toprule
			\textbf{Syntax Element} &
			\textbf{Attribute} &
			\textbf{Description} \\ \midrule
			\textbf{State Machine} &
			&
			\begin{tabular}[c]{@{}l@{}}A finite state machine with a fixed number of [State]s and [Transition]s.\end{tabular} \\ \midrule
			\textbf{State} &
			&
			\begin{tabular}[c]{@{}l@{}}Representation of information of a system at a given point.\end{tabular} \\ \midrule
			\textbf{} &
			name &
			Name of the [State]. \\ \midrule
			\textbf{} &
			type &
			\begin{tabular}[c]{@{}l@{}}Type of a [State]: \textit{Initial},   \textit{Intermediate}, \textit{Final}.\end{tabular} \\ \midrule
			\textbf{Transition} &
			&
			\begin{tabular}[c]{@{}l@{}}A path between two [State]s based on an action.\end{tabular} \\ \midrule
			\textbf{} &
			source &
			\begin{tabular}[c]{@{}l@{}}A [Transition] starts at this [State]. \end{tabular} \\ \midrule
			\textbf{} &
			target &
			\begin{tabular}[c]{@{}l@{}}The [State] where the [Transition] ends.\end{tabular} \\ \midrule
			\textbf{} &
			action &
			\begin{tabular}[c]{@{}l@{}}The [Trigger] that switches the [Transition] from a source to a target [State].\end{tabular} \\ \midrule
			\textbf{Trigger} &
			&
			\begin{tabular}[c]{@{}l@{}}A logical condition for a [Transition] running for a definite period of time.\end{tabular} \\ \midrule
			\textbf{} &
			condition &
			A string holding the condition requirement. \\ \bottomrule
		\end{tabular}
		\label{tab:semantic-mapping-sm}
		\vspace{-2.9em}
	\end{center}
\end{table*}

\subsection{The Method}
\vspace{-0.35em}
The \textit{method} part of the DSL Building Block describes the modelling goal for a business use case, which serves as a guide to a DSL User, and addresses the question "how-to" reach that goal. Modelling goals vary for DSL Users and include representations of target systems and their behaviour, ideas, simulations as well as specific business requirements for a wide variety of projects. An example of a business use-case would be to design a traffic signal system that follows a sequential control logic. Given the variety of domains and projects, there is a need for a comprehensive guide for DSL Users to reach their goals in an effortless manner. The DSL Building Block Developer describes a suitable approach to reach these modelling goals using a sequence of methodical steps, specified textually and using an activity diagram. The inputs and outputs at each step consists of an actual model, parts of a model or trivial and non-trivial business requirements in relation to the composition of the system in consideration. In addition, a list of constraints is defined for the DSL Developer to ensure that the basic, yet critical, conditions for a DSL Building Block, that may otherwise be overlooked, are pre-checked and validated. The methodical steps provide a helping guide to DSL Users in reaching their modelling goals. 

The list of constraints for a finite state machine, which consists of a finite number of [State]s and [Transition]s, are described as follows by the DSL Building Block Developer. \textit{Constraint 1:} All the [State]s described in the State Machine DSL Building Block must be reachable by [Transition]s, except the initial [State] which may or may not be reachable by a [Transition] but is the starting point of the machine. \textit{Constraint 2:} A [State] can have more than one incoming [Transition]s, each from different [State]s. Similarly, a [State] can have more than one outgoing [Transition]s, each to different [State]s. \textit{Constraint 3:} The initial [State] has either zero or one incoming [Transition] from an intermediate [State] and the final [State]s will not have any outgoing [Transition]s. \textit{Constraint 4:} A [Transition] between two [State]s must execute within a defined time frame.

\begin{figure}[h]
	\centering
	\includegraphics[width=\linewidth]{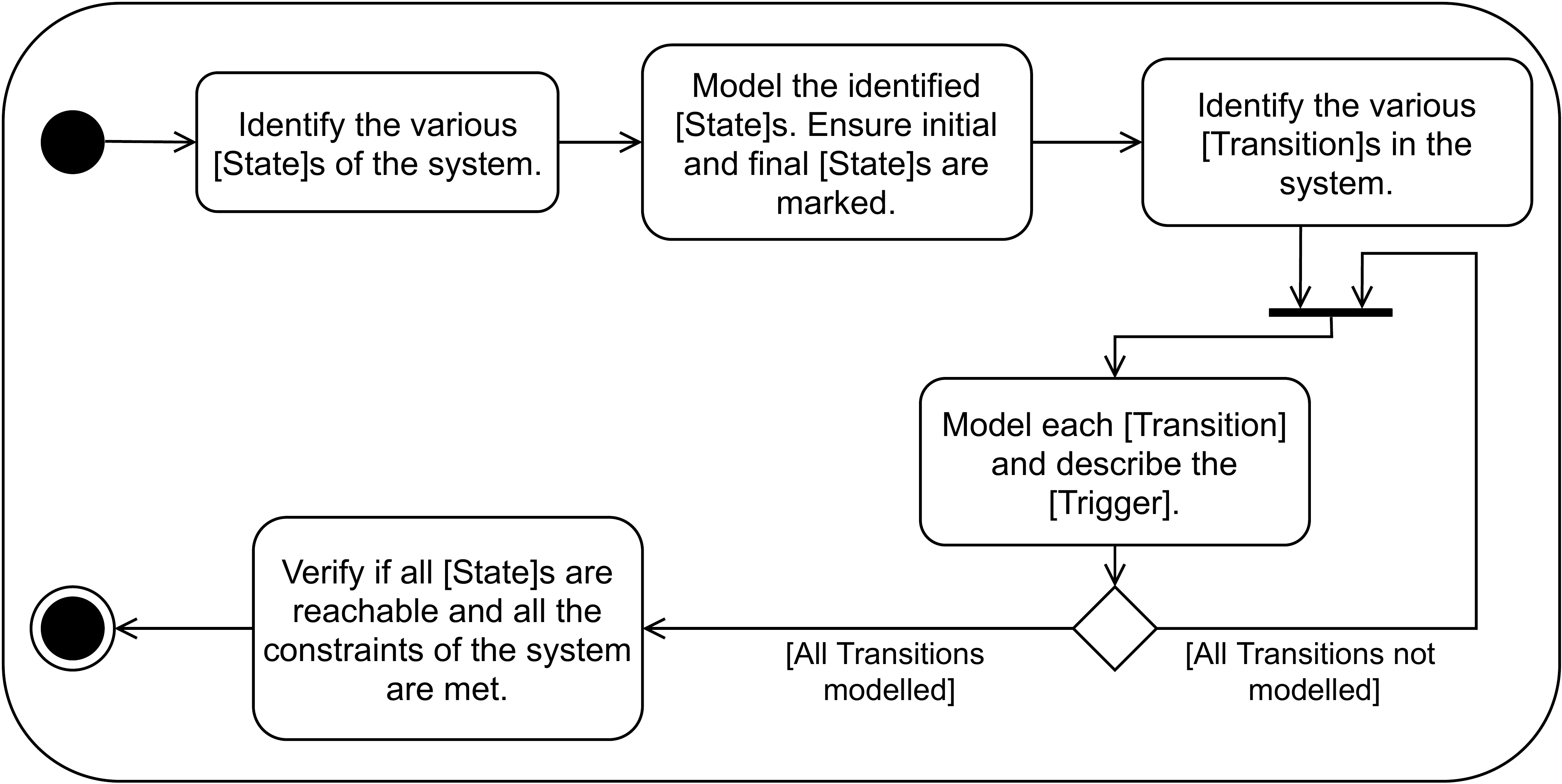}
	\vspace*{-1.4em}
	\caption{A method activity diagram that describes the sequence of methodical steps needed to reach the modelling goal for the state machine DSL Building Block example.}
	\label{fig:dslbb_sm_ad}
	\vspace{-1.25em}
\end{figure}

Fig. \ref{fig:dslbb_sm_ad} describes the steps needed to reach a modelling goal for a state machine making it beneficial for a DSL User to easily understand and follow the process needed to reach their modelling goals.
\vspace{-1em}


\subsection{The Nucleus}
\vspace{-0.5em}
The \textit{nucleus} part of the DSL Building Block consists of various characteristics of model elements for the language, termed \textit{nucleus nuances}, that help provide a coherent UX to users based on context conditions, visual representations, transformations and validation rules. Context conditions are boolean predicates on a language's abstract syntax to checking its consistency and is used to determine if a model is well-formed \cite{HR17}. Visual representations are representations of model elements in the form of icons, colours, appearance, dialogs and its properties in relation to shape, size and opacity. Transformations allow model elements to be automatically instantiated and validation rules enable better error detection with model elements, including checking redundant model elements or misconfigured types that are hard to be detected manually. Certain studies have explored generating graphical syntax for better visual representation \cite{SATG15, NastovP14}. A \textit{nucleus nuance} is a characteristic of a model element describing the model element's intent, motivation and consequences based on reasoning to enhance UX for DSL Users. The usability aspects in building graphical DSLs is often neglected which leads to users struggling in understanding complex DSLs \cite{Mosqueira-ReyA20, RodriguesZBC18}. Therefore, nuances are described by the DSL Building Block Developer for effective usability and visual notation conventions \cite{GreenP96, Blackwell08, 10.1109/TSE.2009.67} in making better design decisions with respect to each individual domain. Each nuance is described with a reasoning as a textual template, that the DSL Developer builds into the DSL Building Block and the DSL. Users can thus easily understand the notations and importance of these nuances for effectively using a graphical DSL. We now list a few nucleus nuances for the state machine example. 

\textit{Nuance 1:} On creation of a [State Machine], an initial [State] is also created automatically on a graphical modelling tool canvas. \textit{Reason}: Users often forget to create or mark the initial state or remove it without marking another initial state during model creation or refactoring.
\vspace{-.2em}

\textit{Nuance 2:} [State]s are oval or circular in shape and [Transition]s are denoted with curved black arrows. \textit{Reason}: Representing different model elements in particular shapes allows for easy visual identification of model elements on the graphical modelling tool.
\vspace{-.2em}

\textit{Nuance 3:} The initial [State] is marked with an \faInfo \space (alphabet i) symbol whereas the final [State]s are marked with an \faFacebook \space (alphabet f) symbol. \textit{Reason}: Visualizations with different symbols prevent users from creating multiple such [State]s or confuse them with other [State]s in a complex system with multiple model elements.
\vspace{-.2em}

\textit{Nuance 4:} All instances of a [State] are filled with a distinct colour, except intermediate [State]s, where two such intermediate [State]s can be filled with the same colour. \textit{Reason}: Visualizations with different colours help distinguish [StateType]s.
\vspace{-.2em}

\textit{Nuance 5:} A [State] without an incoming or outgoing [Transition]s is marked with a red exclamation mark \textcolor{red}{\faExclamation} at its top right corner. \textit{Reason}: Often in complex models, changes in the model leads to the unwanted removal of model elements leading to errors in model. This nuance, thus, helps in validation and error detection.
\vspace{-.2em}

\textit{Nuance 6:} A [Transition] between two [State]s is represented by a curved black line. If a [Transition] does not contain a [Trigger], the link is coloured red along with an exclamation mark \textcolor{red}{\faExclamation}. \textit{Reason}: This nuance also helps in detecting errors and validates certain rules that may be thought of initially as an implied behaviour.
\vspace{-.2em}

\textit{Nuance 7:} Each [State]s can also be marked with any additional relevant icon that represents visually aiding information about the [State]. \textit{Reason}: Complex industrial systems consists of multiple hardware and software resources. Using relevant icons helps identify model elements with ease.
\vspace{-.2em}

\vspace{-.5em}
\section{Extended Example as a Case Study}
\label{cs}
\vspace{-.5em}
A local government of a city is holding an exhibition (expo) inviting industrial manufacturers to foster innovation around mobility and sustainability. To effectively manage traffic during the event, an intelligent traffic management system is needed serving different purposes. For example, one area of the expo calls for a traffic signal serving pedestrians and cars, while another area of the expo needs a traffic signal catering to fully autonomous cars, thus requiring fewer states and transitions and standardised traffic signal lights that the autonomous cars can detect to either proceed or stop. With a state machine DSL Building Block defined earlier, deriving such traffic signal DSL Building Blocks, that also require sequential control logic, becomes significant for re-usability. This reduces repetitive tasks such as defining constraints, states and transitions and defining syntax related to the language during the graphical DSL development.

\begin{figure}[h]
	\centering
	\includegraphics[width=\linewidth]{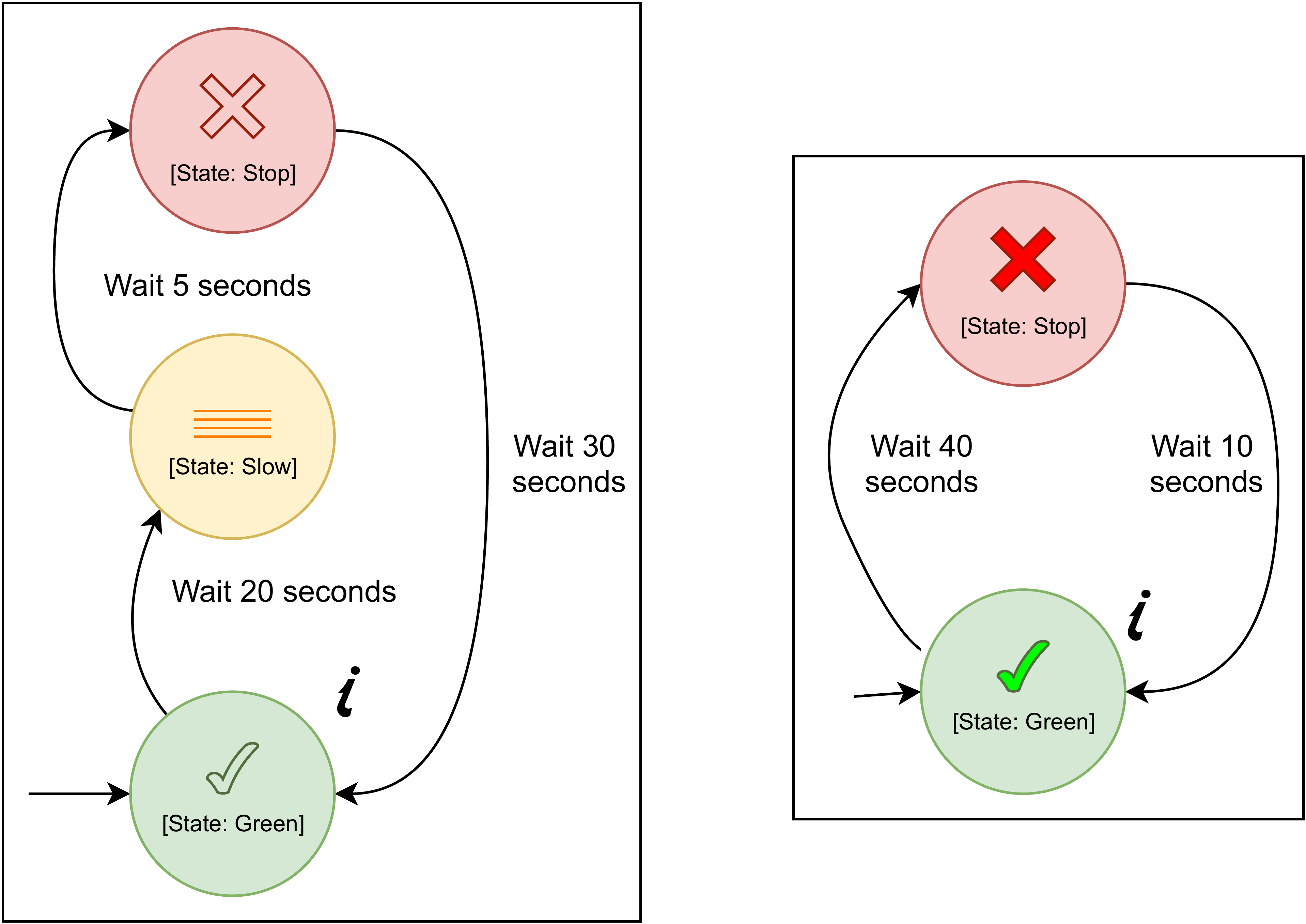}
	\vspace*{-1.9em}
	\caption{A traffic signal model for pedestrians and cars (left) and a model for autonomous cars with fewer states and transitions (right).}
	\label{fig:ts_impl}
	\vspace{-1.85em}
\end{figure}

We extend the example of a state machine to a traffic signal in this case study. A traffic signal DSL Building Block consists of three state instances: {[State: Go]} (Initial State), {[State: Slow]} (Intermediate State), {[State: Stop]} (Intermediate State). The transitions and triggers for these states are shown in Table \ref{tab:traffic-transitions}. The times in the trigger column are filled in by DSL Users when building models, thus simplifying the use of a DSL. The method, language and nucleus parts are adapted by the DSL Building Block Developers for a specific traffic signal and do not influence the state machine DSL Building Block. However, an advantage of exploring different case studies allows common adaptations to also be incorporated directly in a parent DSL Building Block for future re-use, thus allowing quick and continuous improvements. An example of such commonality could be a sensor that is used in all instances of all traffic signal implementation. Finally, nuances help in relating to better visual representations allowing DSL Users to design traffic signal models that are similar to real-world traffic signals as shown Fig. \ref{fig:ts_impl}.

\begin{table}[!h]
	\centering
	\caption{Transitions and trigger conditions for a traffic signal DSL Building Block.}
	\vspace*{-.75em}
	\resizebox{\linewidth}{!}{%
	\begin{tabular}{@{}|l|l|l|l|@{}}
		\toprule
		\textbf{Transition} & \textbf{Source State} & \textbf{Target State} & \textbf{Trigger Condition}    \\ \midrule
		{[}Transition: 1{]} & {[}State: Go{]}       & {[}State: Slow{]}     & "Wait t$_1$ seconds"$^{\mathrm{a}}$   \\ \midrule
		{[}Transition: 2{]} & {[}State:   Slow{]}   & {[}State:   Stop{]}   & "Wait t$_2$ seconds"$^{\mathrm{a}}$  \\ \midrule
		{[}Transition: 3{]} & {[}State:   Stop{]}   & {[}State:   Go{]}     & "Wait t$_3$ seconds"$^{\mathrm{a}}$ \\ \bottomrule
		\multicolumn{4}{l}{$^{\mathrm{a}}$t$_1$, t$_2$, t$_3$ are the times defined by a DSL user when using the DSL.}
	\end{tabular}}
	\label{tab:traffic-transitions}
	\vspace{-1.7em}
\end{table}

We now discuss modifications to the state machine example to support the traffic signal DSL Building Block. The following constraint is added to the method. \textit{Constraint 5:} As a traffic signal is intended to run indefinitely, the final state will not exist. The methodical steps from the state machine example is updated to include the following transitions. {[Transition: 1]}: {[State: Go] -\textgreater [State: Slow]}; {[Transition: 2]}: {[State: Slow] -\textgreater [State: Stop]}; {[Transition: 3]}: {[State: Stop] -\textgreater [State: Go]}. The class diagram representing the abstract syntax is updated to reflect traffic signal specifics, such as two state types: Initial and Intermediate, and three states. The documentation of the model elements described in Table \ref{tab:semantic-mapping-sm} is updated to reflect the changes. Transitions and triggers for this traffic signal example are described in the Table \ref{tab:traffic-transitions}. To further improve visual aspects for the traffic signal, the following nuances are updated or added. \textit{Nuance 2 (updated):} The states must be circular in shape and are arranged vertically in this order (top to bottom): {[State: Stop]}, {[State: Slow]} and {[State: Go]}. \textit{Nuance 8:} {[State: Go]} is filled with green colour and contains a green tick (\textcolor{green}{\faCheck}) icon. \textit{Nuance 9:} {[State: Slow]} is filled with yellow colour and contains an orange (\textcolor{orange}{\faAlignJustify}) icon. Finally, \textit{Nuance 10:} {[State: Stop]} is filled with red colour and contains a red cross (\textcolor{red}{\faTimes}) icon. The reasoning for these nuances is to effectively represent important model elements visually by defining colours, shapes and layouts that are commonly used in real world traffic signals. This case study shows methodical steps, structured documentation and better visual notations of model elements that help DSL Users model systems with a greater degree of confidence and show how parts of a similarly structured graphical DSL can be re-used with minimal adaptations.

\section{Evaluation}
\label{eval}
\vspace{-.4em}
The proposed approach was subject to a qualitative assessment using focus group methods \cite{krueger2009focus}. The purpose of this evaluation was to bring together a group of experienced practitioners and researchers to collectively understand the challenges of the approach, discuss possible solutions and define a systematic graphical DSL developmental approach that would be beneficial for both practitioners and researchers alike. In this section, we describe the evaluation pre-processing, discussions and the results. The evaluation was initially planned for two in-person phases, however due to travel and contact restrictions surrounding COVID-19, the second phase was conducted online using video conferencing tools.
\vspace{-.8em}

\subsection{Participants}
\vspace{-.4em}
To assess the approach, five participants with varying modelling and programming knowledge were chosen. Two participants from the industry were domain modelling experts with 8-12 years experience in developing graphical DSLs. Two participants were researchers from the software engineering domain with limited programming, but 6-12 years experience in software and systems modelling. The final participant was a software developer with limited modelling experience, but 5 years of programming and UX skills. The moderator, a research group manager with experience as a scrum master, was briefed with the proposed approach before each phase.
\vspace{-.9em}
\subsection{Phase 1}
\vspace{-.6em}
Phase 1 was held in-person. The focus group was guided by the moderator who prepared a list of questions and activities on the proposed approach with the discussion lasting two hours.

\textbf{Pre-processing.} The moderator was given three weeks to prepare the content of the discussion, the scripts and the technical setup. The outcome of this phase was to provide a first look into the problems faced by domain experts in building graphical DSLs for industrial projects, and to foster lively discussions and feedback on the proposed approach. The moderator proposed the prepared set of questions and activities, including whiteboard discussions, and took hand-written notes while guiding the discussion.

\textbf{Discussion and results.} The discussion was held over three stages: introduction, main stage and the follow-up phase. In the introduction stage, the moderator introduced the participants and presented the elementary question: (1) \textit{What is the single most difficult challenge when designing DSLs?} The practitioners and researchers unanimously agreed on the need to have proper guidelines in developing graphical DSLs. In the main stage, the following questions were asked: (2) \textit{Which steps in designing graphical DSLs consumes the most of your time?} (3) \textit{Is the usage of visual notations beneficial when using a graphical DSL?} In this stage, there were mixed opinions. While practitioners favoured UX as being more beneficial, the researchers said re-usability helps in reducing time-consuming tasks. Then, the difference between the language and nucleus parts was discussed as certain nuances could be part of the syntax of the language. However, one researcher opined that segregating the nuances and classifying them with reasoning would be more structured for developers. The software developer advocated the use of common visual notations and representations making DSL elements easier for any user to understand. In the follow-up phase, all participants believed more clarifications were needed for phase two as some issues of this approach were insufficiently addressed.

\vspace{-.6em}
\subsection{Phase 2}
\vspace{-.5em}
Phase 2, initially planned for in-person, was conducted online, three months later, due to restrictions surrounding COVID-19 with the same participants and moderator. Building on the experience in the first phase and being updated with an improved version of the approach, the moderator framed a diverse set of questions and activities for this phase.

\textbf{Pre-processing.} The moderator prepared the content of the discussion and set up the audio and video conferencing in Microsoft Teams \cite{MSTeams}. The whiteboard discussion was held over Conceptboard \cite{Conceptboard}. The outcome of this phase was to reach a systematic approach beneficial for both users and developers of graphical DSLs.

\textbf{Discussion and results.} The discussion was held over three stages: introduction, main stage and the follow-up phase. In the introduction, the moderator presented the approach and the initial question: (1) \textit{What strengths or weaknesses does this improved approach carry?} This was followed by a creative discussion in the main stage which included virtual whiteboard discussions. In this stage, participants came up with an approach to define various constraints in the method part and reasoning for each nuance based on the state machine example. At times the audio and video distorted, however, the moderator noted that there was no real disruption in the outcome as all topics were well articulated. The inability to use an offline whiteboard was a challenge for participants as more time was needed here than planned. In the follow-up phase, participants were asked to describe and rate the further improved approach and list prospective work. The participants agreed this final systematic approach is beneficial for both developers and users as well as practitioners and researchers. The final approach presented in this paper is a three layered developmental approach incorporating various suggestions and feedback based on both focus group discussions.

\vspace{-.45em}
\section{Discussion}
\label{discussion}
\vspace{-.35em}

The presented methodology enables DSL Developers to systematically develop graphical DSLs. We achieve re-usability of common aspects during graphical DSL development by separating the concerns of industrial DSL engineering along three different levels relating to different skill sets and activities. The approach, DSL Building Blocks, assists DSL Building Block Developers in defining constraints and methodical steps intended to help DSL Users achieve their modelling goals. The business requirements for each project specific DSL is gathered by DSL Building Block Developers in consultation with DSL Users. The developer then extracts commonalities from requirements in the form of a DSL Building Block definition. Project specific use-cases of DSL Building Blocks can extend and adapt the method, language and nucleus parts, thereby providing adequate flexibility to domain experts in realizing problems of a specific domain. Furthermore, visual representation of various DSL elements is beneficial for DSL Users in simplifying their DSL modelling experience, as users can relate such elements to real world domain specific examples. The combination of providing methodical steps to reach a modelling goal and the focus on UX for DSL Users has been addressed with our approach. Our approach presents a structured graphical DSL development process including adaptation and continuous integration of DSL requirements from DSL Users, who closely interact and suggest feedback to DSL Building Block Developers. Our approach to DSL engineering is limited to graphical DSLs, as large corporations such as Siemens, mostly focus on visualization concepts for representing DSLs. We note that graphical modelling tools such as MetaEdit+ \cite{Tolvanen06} and MagicDraw \cite{Neuendorf06} have different technical capabilities, which therefore poses a challenge towards adopting this methodology seamlessly across all tools. Other language engineering tools such as MPS \cite{VolterV10}, Spoofax \cite{WKV14}, and Melange \cite{DCB+15} provide certain means for language composition and customization, but fail to provide methods for systematic reuse for similarly structured DSLs. While our focus group evaluation was limited to a few practitioners and researchers, we are currently building an extensive survey to gather ways on improving UX for domain experts within Siemens. As part of the ongoing research, we plan to categorize and structure various nuances allowing for them to be easily analyzed to make it more machine-processable and accessible to automation. Further, we plan to work on challenges to introduce inheritance support for multiple DSL Building Blocks in enabling re-use support across a wider variety of domains not structurally related to each other.

Overall, our approach builds upon our experience of developing graphical DSLs for various industry projects as well as the qualitative evaluation of the approach using focus groups that included experienced practitioners and researchers. In this vision, the need to have a proper set of guidelines and documentation is important. We, therefore, propose this systematic approach that not only fosters re-use of language parts for DSL Developers, but also focuses on visual representation of model elements that help DSL Users understand and use graphical DSLs with simplicity. We are unaware of any other similar methodologies for the development of graphical DSLs that helps realize this vision.

\vspace{-.5em}
\section{Conclusions}
\label{con}
\vspace{-.5em}
We have presented a systematic approach for developing graphical DSLs, DSL Building Blocks, through separation of concerns of industrial DSL engineering among developers and users with different skill sets and activities. This approach solves challenges related to re-usability of common DSL elements for developers and is intended for a better user experience for domain experts who possess different skill sets in DSL language use and engineering. While our approach is currently limited to graphical DSLs, they greatly facilitate the re-use of DSL parts and provide robust guidance, documentation and effective visual representations to users helping them achieve modelling goals with simplicity. Continuous feedback by users to developers help in constantly adapting and improving the DSL Building Blocks and subsequent DSLs. This fosters the adoption of the systematic development of graphical DSLs among developers and narrows the gap between practitioners and researchers. As part of our ongoing research, we are building an extensive feedback survey to collect experiences from users in improving our methodology. We also plan to apply our approach to different domains by introducing inheritance support between DSL Building Blocks and integrate further language definition dimensions.

\section*{Acknowledgement}
\vspace{-.5em}
The authors would like to thank Ambra Calà and Jérôme Pfeiffer for their inputs during the course of writing this paper.

\bibliographystyle{ieeetran}
\bibliography{bib,se,aw}
	
\end{document}